\documentclass[superscriptaddress, twocolumn, amsmath, amssymb, aps, floatfix, pra]{revtex4-2}

\usepackage{graphicx}
\usepackage{bm}
\usepackage{mathrsfs}
\usepackage{physics}
\usepackage{xcolor}
\usepackage{amsmath, amssymb, mathtools, amsthm}
\usepackage[colorlinks=true]{hyperref}
\usepackage{orcidlink}

\begin{document}

\title{Synthesizing superoscillations with just two frequencies}

\author{Denys I. Bondar\orcidlink{0000-0002-3626-4804}}
\email{dbondar@tulane.edu}
\affiliation{Tulane University, New Orleans, Louisiana 70118, United States}

\author{Diyar Talbayev\orcidlink{0000-0003-3537-1656}}
\email{dtalbaye@tulane.edu}
\affiliation{Tulane University, New Orleans, Louisiana 70118, United States}

\date{\today}

\begin{abstract}
Superoscillations, band-limited signals that locally oscillate faster than their highest Fourier component, have recently enabled superspectroscopy and super-sensing. Previous experiments [\href{https://doi.org/10.1103/PhysRevLett.131.153803}{Phys. Rev. Lett. 131, 153803 (2023)} and \href{https://doi.org/10.1063/5.0271556}{APL Photonics 10, 086107 (2025)}] relied on combining four quasi-sinusoidal harmonics to generate temporal superoscillations. Here, we show that just two harmonics are sufficient to produce superoscillations of comparable quality, as quantified by the local frequency. By reanalyzing prior experimental data with two harmonics ($0.5$ and $0.6$~THz) and performing new experiments with $0.9$ and $1$~THz harmonics, we demonstrate a twofold enhancement in local frequency relative to the highest frequency component---matching the enhancement previously achieved with four harmonics. We provide a simple analytical explanation for the bichromatic superoscillations based on near-complete destructive interference, and identify an optimal frequency separation of approximately $10\%$ that balances the superoscillatory frequency enhancement against signal amplitude. This simplification substantially lowers the experimental barrier to implementing superoscillation-based technology.
\end{abstract}

\maketitle

\section{Introduction}

A superoscillation is a band-limited signal that, within a localized spatial or temporal region, oscillates faster than its highest Fourier component~\cite{berry_roadmap_2019, rogers_realising_2020}. Although such behavior may appear counterintuitive, superoscillations arise naturally from the near-complete destructive interference of the constituent frequency components~\cite{berry_faster_1994}. The concept was introduced in Ref.~\cite{berry_faster_1994} in connection with weak measurements~\cite{aharonov_time_1964, tamir_introduction_2013}, and has since developed into an active area of research spanning mathematics and physics~\cite{berry_roadmap_2019, jordan_superoscillations_2025}.

In the spatial domain, superoscillations have been realized experimentally and applied to sub-wavelength imaging and super-resolution microscopy~\cite{rogers_super-oscillatory_2012, Rogers_2013}. In the temporal domain, however, progress has been slower. Radio-frequency superoscillations have been demonstrated~\cite{wong_temporal_2011, wong_superoscillatory_2012}, and superoscillating laser pulse envelopes have been constructed at optical frequencies~\cite{eliezer_breaking_2017, eliezer_experimental_2018}. Sub-terahertz superoscillations have also been observed in coherent acoustic phonon dynamics~\cite{brehm_temporal_2020}.

We recently achieved the synthesis of time-domain optical superoscillations at terahertz (THz) frequencies~\cite{mccaul_superoscillations_2023}. The obtained superoscillations were used to demonstrate ``superspectroscopy''---an enhancement of nearly an order of magnitude in the linear spectroscopic sensitivity to materials whose resonances lie outside the frequency band of the source waveforms. More recently, this approach was extended to achieve a 100-fold contrast enhancement in THz time-domain spectroscopy~\cite{peng_super-sensing_2025}. Beyond these applications, superoscillatory signals have also been shown to survive reconstruction even when buried deeply in noise~\cite{white_reconstructing_2024}, underscoring their practical viability.

These prior demonstrations, however, employed four harmonics, which requires a complex experimental setup involving multiple THz generation channels, beamsplitters, and independent delay stages. A natural question is whether superoscillations of comparable quality can be produced with fewer harmonics. In this article, we affirmatively answer this question: Just two harmonics are sufficient to generate superoscillations with the same twofold local frequency enhancement achieved using four harmonics. We support this conclusion with both a reanalysis of existing experimental data and new experiments, and provide a simple analytical argument explaining why bichromatic superoscillations arise from destructive interference between two closely spaced frequencies.

\section{Results}

In Ref.~\cite{mccaul_superoscillations_2023}, we put forth an experimental method to synthesize a superoscillating optical field using a set of time-limited $E_j(t)$ waveforms with different central frequencies $f_j$,
\begin{align}
    E(t) = \sum_{j=1}^N E_j(t-\tau_j),
\end{align}
by finding the time delays $\tau_j$ that minimize the intensity of $E(t)$ within a time window $t\in [-T_{\rm SO},T_{\rm SO}]$ via minimization of the cost function
\begin{align}
\label{eq:costfunction}
    I(\tau_1, \ldots, \tau_N) = \int_{-T_{\rm SO}}^{T_{\rm SO}} \Bigg[ \sum_{j=1}^N E_j(t - \tau_j) \Bigg]^2 {\rm d}t. 
\end{align}
Note that $T_{\rm SO}$ is typically chosen as half the period of the highest frequency component, $T_{\rm SO}  = \pi / \max\{ f_j \}$.

In Ref.~\cite{mccaul_superoscillations_2023}, we experimentally used four ($N=4$) near-sinusoidal waveforms $E_{j}(t)$ with corresponding frequencies $f_j= 0.52,\, 0.63,\, 0.72,\, 0.82$~THz. 

The purpose of this brief article is to show that it is sufficient to use only two harmonics to generate superoscillations of the same quality (as characterized by local frequency) as using four harmonics. 

The precise enhancement in frequency provided by the superoscillation is characterized by the local frequency. This is defined in analogy with the local wavenumber used to characterize spatial superoscillations~\cite{Rogers_2013, stern_light_2024}, and is the gradient of the phase, 
\begin{align}
    f_{\rm loc}=\frac{1}{2\pi}\frac{{\rm d} \phi}{{\rm d}t}.    
\end{align}
This is obtained by Hilbert transforming the real-valued signal $E(t)$ into an analytic form~\cite{smith2007mathematics} $\hat{E}(t)$, from which the field phase $\phi(t) = \arg\hat{E}(t)$ is extracted. 

To demonstrate that $N=2$ is indeed all that is needed, we first reanalyzed the experimental data from Fig.~3 of Ref.~\cite{mccaul_superoscillations_2023} using two harmonics, $f_1 = 0.5$~THz and $f_2 = 0.6$~THz. In experiments, quasi-sinusoidal THz harmonics are emitted via optical rectification of focused, 160-femtosecond laser pulses with 1030 nm center wavelength in periodically poled lithium niobate (PPLN) \cite{lee_generation_2000, lee_temperature_2000, weiss_tuning_2001, lhuillier_generation_2007, lhuillier_generation_2007-1}. All PPLN chips were 5 mm long with 0.5 mm thickness. The poling periods for the 0.5 THz and 0.6 THz PPLN emitters were 208 $\mu$m and 173 $\mu$m, respectively. The THz waveforms were detected via free-space electro-optic sampling in a ZnTe crystal \cite{lin_measurement_2018}. We performed the minimization of Eq.~\eqref{eq:costfunction} to find the time delays $\tau$ that produce superoscillations when the two harmonics are added. The results are shown in Fig.~\ref{fig_bichromatic_so_old_data} for three different values of $\tau$ (the minimization typically produces many local minima of the cost function~\eqref{eq:costfunction}, each with its own $\tau$ value). In particular, the local frequencies plotted in Figs.~\ref{fig_bichromatic_so_old_data}(g, h, i) show a twofold increase relative to the highest frequency component within the superoscillating window---the same increase as in Ref.~\cite{mccaul_superoscillations_2023}, where $N=4$ harmonics were used. 

Furthermore, we performed a new set of experiments using $0.9$ and $1$~THz harmonics, for which we used PPLN chips with 115 $\mu$m and 103 $\mu$m poling periods. Here, we first measure each harmonic separately, then synthesize superoscillations computationally by finding the time delays $\tau$. We then implement the time delays $\tau$ in the experiment and measure the combined waveform $E(t)$. Figure~\ref{fig_theory_vs_experiment_so} shows the combined computational and experimental waveforms for three values of $\tau$. The figure experimentally confirms the theoretical prediction that two harmonics are sufficient for producing superoscillations.  

As anticipated, Figs.~\ref{fig_theory_vs_experiment_so}(g, h, i) show that for most of the duration of the combined waveform, its local frequency is approximately equal to its highest frequency component. Within the region of minimization, however, the combined waveform's $f_{\rm loc}$ deviates sharply, experiencing an approximately twofold increase relative to the highest frequency component. 

\begin{figure}
    \centering
    \includegraphics[width=1\linewidth]{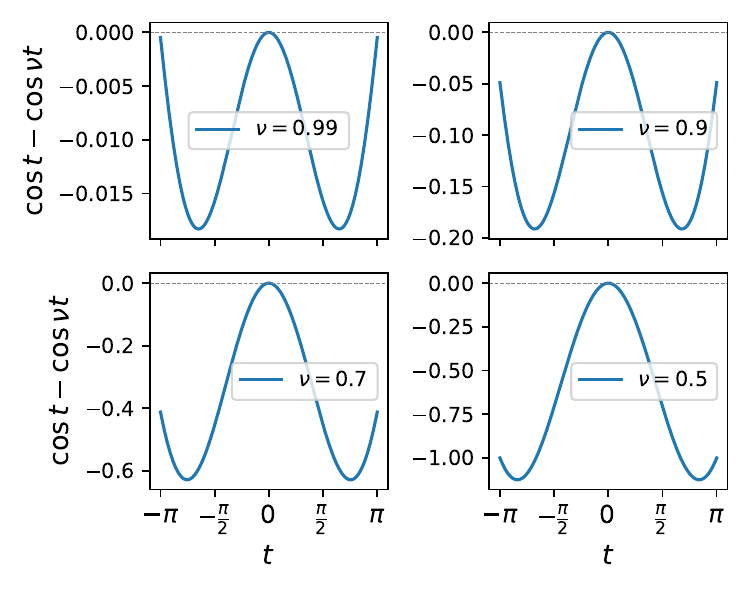}
    \caption{An illustration of bichromatic superoscillations using~\eqref{EqSimpleF} for several values of $\nu$. The closer $\nu$ approaches $1$, the closer the function comes to exhibiting two oscillations within the superoscillating region. When $\nu = 1$, there is complete destructive interference.}
    \label{fig_explanation_bichromatic_so}
\end{figure}

A simple example explains why two harmonics, when destructively interfering, produce superoscillations. Consider the function 
\begin{align}\label{EqSimpleF}
    f(\nu; t) = \cos t - \cos \nu t, \qquad \nu < 1,
\end{align}
illustrated in Fig.~\ref{fig_explanation_bichromatic_so}. The closer $\nu$ is to $1$, the closer the local frequency approaches a twofold enhancement. However, there is a trade-off: the amplitude of the superoscillation grows weaker as $\nu$ approaches $1$. Figure~\ref{fig_explanation_bichromatic_so} suggests that the optimal trade-off for generating superoscillations is to use two frequencies that differ by approximately $10\%$. In this case, the local frequency is doubled while the superoscillating region retains sufficient intensity for experimental detection. This is precisely what is observed in Figs.~\ref{fig_bichromatic_so_old_data} and~\ref{fig_theory_vs_experiment_so}.

\section{Conclusions}

We have demonstrated that superoscillations of the same quality as those produced with four harmonics can be synthesized using only two. This finding has important practical implications. Four-color superoscillations have recently been shown to enable super-sensing---a 100-fold contrast enhancement in linear THz time-domain spectroscopy~\cite{peng_super-sensing_2025}. Since bichromatic superoscillations achieve nearly the same local frequency as their four-color counterparts~\cite{mccaul_superoscillations_2023}, we expect that the super-sensing advantage can be retained with a significantly simpler experimental setup. Reducing the number of required harmonics from four to two eliminates the need for multiple THz generation channels and associated optical components, making superoscillation-based spectroscopy and sensing considerably more accessible.

\acknowledgments
The authors acknowledge the very generous support of the W. M. Keck Foundation. D.I.B. was supported by Army Research Office (ARO) (grant W911NF-23-1-0288; program manager Dr.~James Joseph). The views and conclusions contained in this document are those of the authors and should not be interpreted as representing the official policies, either expressed or implied, of ARO or the U.S. Government. The U.S. Government is authorized to reproduce and distribute reprints for Government purposes notwithstanding any copyright notation herein.

\section*{Data Availability}
All codes and data used in this study can be found in Ref.~\cite{your_code_ref}.

\begin{figure*}
    \centering
    \includegraphics[width=0.9\textwidth]{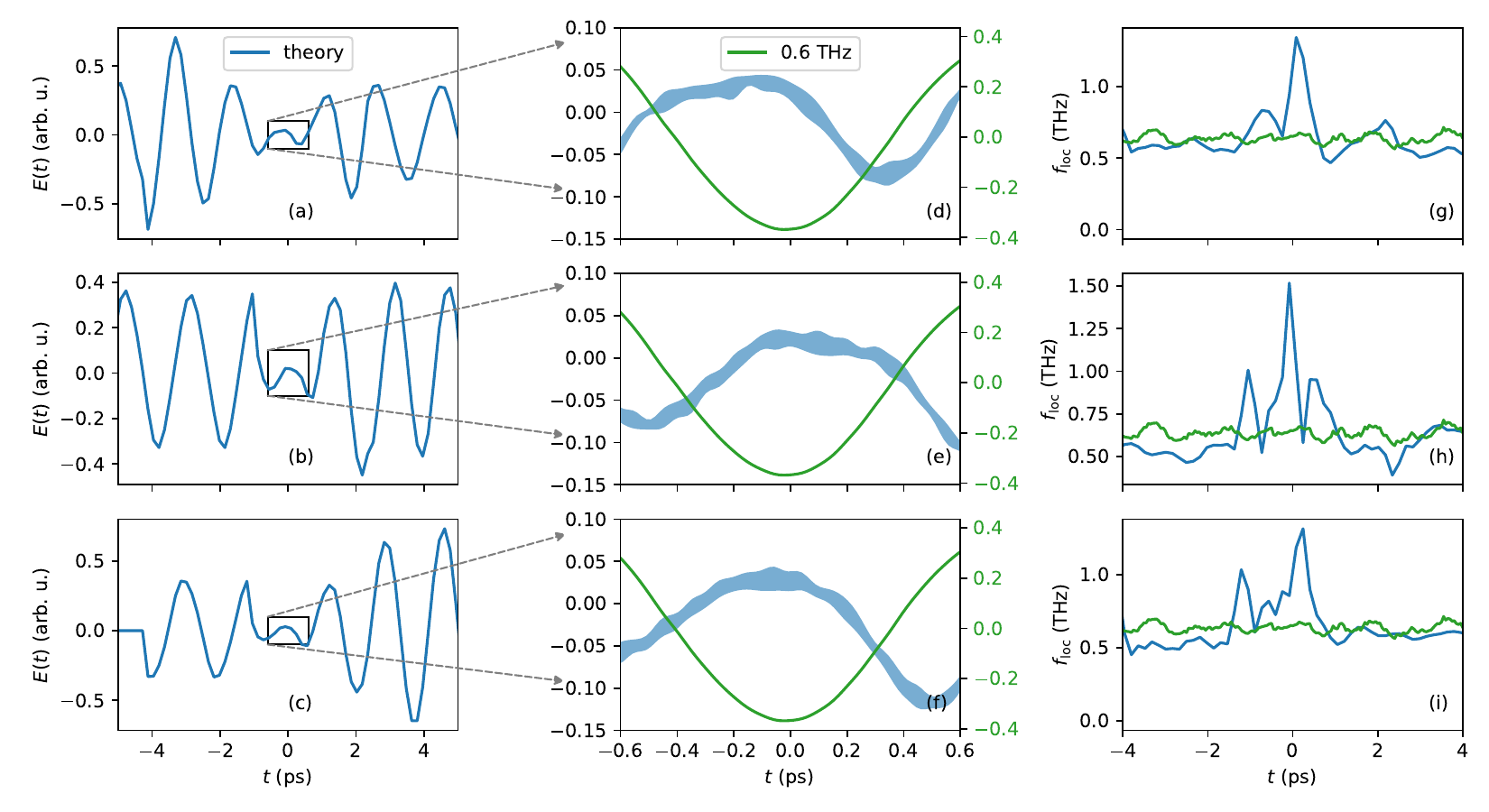}
    \caption{Three example bichromatic superoscillations obtained via the minimization of Eq.~\eqref{eq:costfunction} (using $0.5$ and $0.6$~THz harmonics) by reanalyzing experimental data from Fig.~3 of Ref.~\cite{mccaul_superoscillations_2023}. Left column (a, b, c) shows the predicted field over the full measurement window. The middle panels (d, e, f) highlight the window in which the superoscillation occurs, with comparison to the highest frequency component waveform. The shaded blue area shows the theoretical prediction with the 95\% confidence limit. The right column (g, h, i) plots the local frequencies of both the total waveform and its highest frequency component. Outside the window of minimization these are roughly equivalent, but within this window the combined waveform's local frequency increases sharply compared to that of the highest frequency component. This figure was generated by the code~\footnote{\url{https://github.com/dibondar/bicolor-superoscillations/blob/main/get_superoscilations_from_03-11-2022_data.ipynb}}.}
    \label{fig_bichromatic_so_old_data}
    \includegraphics[width=0.9\textwidth]{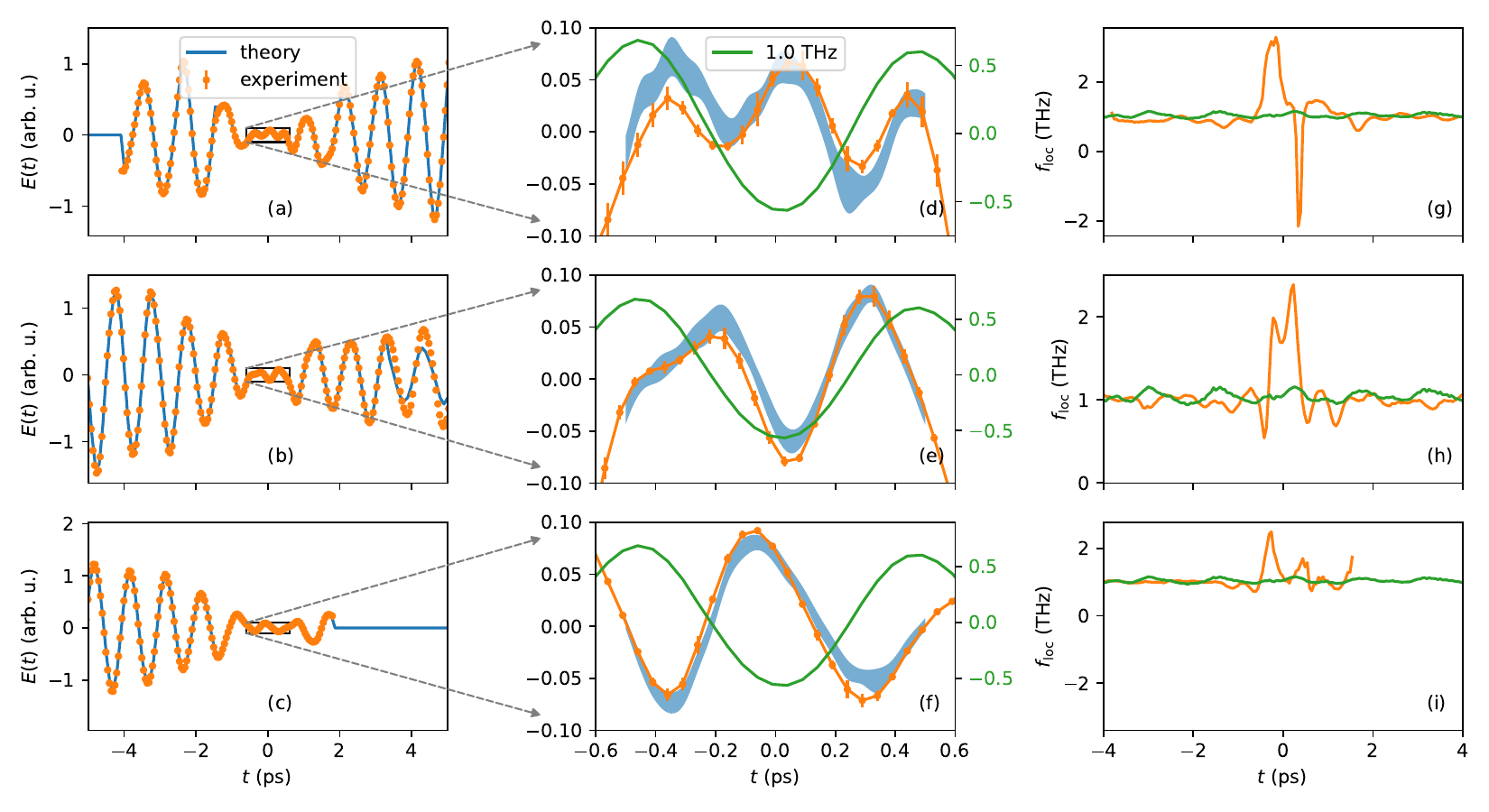}
    \caption{Three example bichromatic superoscillations obtained via the minimization of Eq.~\eqref{eq:costfunction} using $0.9$ and $1$~THz harmonics. Left column (a, b, c) shows both the predicted and observed field over the full measurement window. The middle panels (d, e, f) highlight the window in which the superoscillation occurs, with comparison to the highest frequency component waveform. The experimental data is in excellent agreement with the theoretical 95\% confidence limit (shaded blue area). The right column (g, h, i) plots the local frequencies of both the total waveform and its highest frequency component. Outside the window of minimization these are roughly equivalent, but within this window the combined waveform's local frequency increases sharply compared to that of the highest frequency component. This figure was generated by the code~\footnote{\url{https://github.com/dibondar/bicolor-superoscillations/blob/main/get_superoscilations_from_06-09-2025_data.ipynb}}.}
    \label{fig_theory_vs_experiment_so}
\end{figure*}

\bibliography{references}

\begin{thebibliography}{25}%
\makeatletter
\providecommand \@ifxundefined [1]{%
 \@ifx{#1\undefined}
}%
\providecommand \@ifnum [1]{%
 \ifnum #1\expandafter \@firstoftwo
 \else \expandafter \@secondoftwo
 \fi
}%
\providecommand \@ifx [1]{%
 \ifx #1\expandafter \@firstoftwo
 \else \expandafter \@secondoftwo
 \fi
}%
\providecommand \natexlab [1]{#1}%
\providecommand \enquote  [1]{``#1''}%
\providecommand \bibnamefont  [1]{#1}%
\providecommand \bibfnamefont [1]{#1}%
\providecommand \citenamefont [1]{#1}%
\providecommand \href@noop [0]{\@secondoftwo}%
\providecommand \href [0]{\begingroup \@sanitize@url \@href}%
\providecommand \@href[1]{\@@startlink{#1}\@@href}%
\providecommand \@@href[1]{\endgroup#1\@@endlink}%
\providecommand \@sanitize@url [0]{\catcode `\\12\catcode `\$12\catcode `\&12\catcode `\#12\catcode `\^12\catcode `\_12\catcode `\%12\relax}%
\providecommand \@@startlink[1]{}%
\providecommand \@@endlink[0]{}%
\providecommand \url  [0]{\begingroup\@sanitize@url \@url }%
\providecommand \@url [1]{\endgroup\@href {#1}{\urlprefix }}%
\providecommand \urlprefix  [0]{URL }%
\providecommand \Eprint [0]{\href }%
\providecommand \doibase [0]{https://doi.org/}%
\providecommand \selectlanguage [0]{\@gobble}%
\providecommand \bibinfo  [0]{\@secondoftwo}%
\providecommand \bibfield  [0]{\@secondoftwo}%
\providecommand \translation [1]{[#1]}%
\providecommand \BibitemOpen [0]{}%
\providecommand \bibitemStop [0]{}%
\providecommand \bibitemNoStop [0]{.\EOS\space}%
\providecommand \EOS [0]{\spacefactor3000\relax}%
\providecommand \BibitemShut  [1]{\csname bibitem#1\endcsname}%
\let\auto@bib@innerbib\@empty
\bibitem [{\citenamefont {Berry}\ \emph {et~al.}(2019)\citenamefont {Berry}, \citenamefont {Zheludev}, \citenamefont {Aharonov}, \citenamefont {Colombo}, \citenamefont {Sabadini}, \citenamefont {Struppa}, \citenamefont {Tollaksen}, \citenamefont {Rogers}, \citenamefont {Qin}, \citenamefont {Hong} \emph {et~al.}}]{berry_roadmap_2019}%
  \BibitemOpen
  \bibfield  {author} {\bibinfo {author} {\bibfnamefont {M.}~\bibnamefont {Berry}}, \bibinfo {author} {\bibfnamefont {N.}~\bibnamefont {Zheludev}}, \bibinfo {author} {\bibfnamefont {Y.}~\bibnamefont {Aharonov}}, \bibinfo {author} {\bibfnamefont {F.}~\bibnamefont {Colombo}}, \bibinfo {author} {\bibfnamefont {I.}~\bibnamefont {Sabadini}}, \bibinfo {author} {\bibfnamefont {D.~C.}\ \bibnamefont {Struppa}}, \bibinfo {author} {\bibfnamefont {J.}~\bibnamefont {Tollaksen}}, \bibinfo {author} {\bibfnamefont {E.~T.~F.}\ \bibnamefont {Rogers}}, \bibinfo {author} {\bibfnamefont {F.}~\bibnamefont {Qin}}, \bibinfo {author} {\bibfnamefont {M.}~\bibnamefont {Hong}}, \emph {et~al.},\ }\bibfield  {title} {\bibinfo {title} {Roadmap on superoscillations},\ }\href {https://doi.org/10.1088/2040-8986/ab0191} {\bibfield  {journal} {\bibinfo  {journal} {J. Opt.}\ }\textbf {\bibinfo {volume} {21}},\ \bibinfo {pages} {053002} (\bibinfo {year} {2019})}\BibitemShut {NoStop}%
\bibitem [{\citenamefont {Rogers}\ and\ \citenamefont {Rogers}(2020)}]{rogers_realising_2020}%
  \BibitemOpen
  \bibfield  {author} {\bibinfo {author} {\bibfnamefont {K.~S.}\ \bibnamefont {Rogers}}\ and\ \bibinfo {author} {\bibfnamefont {E.~T.~F.}\ \bibnamefont {Rogers}},\ }\bibfield  {title} {\bibinfo {title} {Realising superoscillations: A review of mathematical tools and their application},\ }\href {https://doi.org/10.1088/2515-7647/aba5a7} {\bibfield  {journal} {\bibinfo  {journal} {J. Phys.: Photonics}\ }\textbf {\bibinfo {volume} {2}},\ \bibinfo {pages} {042004} (\bibinfo {year} {2020})}\BibitemShut {NoStop}%
\bibitem [{\citenamefont {Berry}(1994)}]{berry_faster_1994}%
  \BibitemOpen
  \bibfield  {author} {\bibinfo {author} {\bibfnamefont {M.~V.}\ \bibnamefont {Berry}},\ }\bibfield  {title} {\bibinfo {title} {Faster than {Fourier}},\ }in\ \href@noop {} {\emph {\bibinfo {booktitle} {Quantum Coherence and Reality; in Celebration of the 60th Birthday of {Yakir} {Aharonov}}}},\ \bibinfo {editor} {edited by\ \bibinfo {editor} {\bibfnamefont {J.~S.}\ \bibnamefont {Anandan}}\ and\ \bibinfo {editor} {\bibfnamefont {J.~L.}\ \bibnamefont {Safko}}}\ (\bibinfo  {publisher} {World Scientific},\ \bibinfo {address} {Singapore},\ \bibinfo {year} {1994})\ pp.\ \bibinfo {pages} {55--65}\BibitemShut {NoStop}%
\bibitem [{\citenamefont {Aharonov}\ \emph {et~al.}(1964)\citenamefont {Aharonov}, \citenamefont {Bergmann},\ and\ \citenamefont {Lebowitz}}]{aharonov_time_1964}%
  \BibitemOpen
  \bibfield  {author} {\bibinfo {author} {\bibfnamefont {Y.}~\bibnamefont {Aharonov}}, \bibinfo {author} {\bibfnamefont {P.~G.}\ \bibnamefont {Bergmann}},\ and\ \bibinfo {author} {\bibfnamefont {J.~L.}\ \bibnamefont {Lebowitz}},\ }\bibfield  {title} {\bibinfo {title} {Time symmetry in the quantum process of measurement},\ }\href {https://doi.org/10.1103/PhysRev.134.B1410} {\bibfield  {journal} {\bibinfo  {journal} {Phys. Rev.}\ }\textbf {\bibinfo {volume} {134}},\ \bibinfo {pages} {B1410} (\bibinfo {year} {1964})}\BibitemShut {NoStop}%
\bibitem [{\citenamefont {Tamir}\ and\ \citenamefont {Cohen}(2013)}]{tamir_introduction_2013}%
  \BibitemOpen
  \bibfield  {author} {\bibinfo {author} {\bibfnamefont {B.}~\bibnamefont {Tamir}}\ and\ \bibinfo {author} {\bibfnamefont {E.}~\bibnamefont {Cohen}},\ }\bibfield  {title} {\bibinfo {title} {Introduction to weak measurements and weak values},\ }\href {https://doi.org/10.12743/quanta.v2i1.14} {\bibfield  {journal} {\bibinfo  {journal} {Quanta}\ }\textbf {\bibinfo {volume} {2}},\ \bibinfo {pages} {7} (\bibinfo {year} {2013})}\BibitemShut {NoStop}%
\bibitem [{\citenamefont {Jordan}\ \emph {et~al.}(2025)\citenamefont {Jordan}, \citenamefont {Howell}, \citenamefont {Vamivakas},\ and\ \citenamefont {Karimi}}]{jordan_superoscillations_2025}%
  \BibitemOpen
  \bibfield  {author} {\bibinfo {author} {\bibfnamefont {A.~N.}\ \bibnamefont {Jordan}}, \bibinfo {author} {\bibfnamefont {J.~C.}\ \bibnamefont {Howell}}, \bibinfo {author} {\bibfnamefont {N.}~\bibnamefont {Vamivakas}},\ and\ \bibinfo {author} {\bibfnamefont {E.}~\bibnamefont {Karimi}},\ }\bibfield  {title} {\bibinfo {title} {Superoscillations and physical applications},\ }in\ \href {https://doi.org/10.1007/978-3-0348-0692-3_101-1} {\emph {\bibinfo {booktitle} {Operator Theory}}},\ \bibinfo {editor} {edited by\ \bibinfo {editor} {\bibfnamefont {D.}~\bibnamefont {Alpay}}, \bibinfo {editor} {\bibfnamefont {I.}~\bibnamefont {Sabadini}},\ and\ \bibinfo {editor} {\bibfnamefont {F.}~\bibnamefont {Colombo}}}\ (\bibinfo  {publisher} {Springer, Basel},\ \bibinfo {year} {2025})\ \bibinfo {note} {arXiv:2505.22925}\BibitemShut {NoStop}%
\bibitem [{\citenamefont {Rogers}\ \emph {et~al.}(2012)\citenamefont {Rogers}, \citenamefont {Lindberg}, \citenamefont {Roy}, \citenamefont {Savo}, \citenamefont {Chad}, \citenamefont {Dennis},\ and\ \citenamefont {Zheludev}}]{rogers_super-oscillatory_2012}%
  \BibitemOpen
  \bibfield  {author} {\bibinfo {author} {\bibfnamefont {E.~T.~F.}\ \bibnamefont {Rogers}}, \bibinfo {author} {\bibfnamefont {J.}~\bibnamefont {Lindberg}}, \bibinfo {author} {\bibfnamefont {T.}~\bibnamefont {Roy}}, \bibinfo {author} {\bibfnamefont {S.}~\bibnamefont {Savo}}, \bibinfo {author} {\bibfnamefont {J.~E.}\ \bibnamefont {Chad}}, \bibinfo {author} {\bibfnamefont {M.~R.}\ \bibnamefont {Dennis}},\ and\ \bibinfo {author} {\bibfnamefont {N.~I.}\ \bibnamefont {Zheludev}},\ }\bibfield  {title} {\bibinfo {title} {A super-oscillatory lens optical microscope for subwavelength imaging},\ }\href {https://doi.org/10.1038/nmat3280} {\bibfield  {journal} {\bibinfo  {journal} {Nat. Mater.}\ }\textbf {\bibinfo {volume} {11}},\ \bibinfo {pages} {432} (\bibinfo {year} {2012})}\BibitemShut {NoStop}%
\bibitem [{\citenamefont {Rogers}\ and\ \citenamefont {Zheludev}(2013)}]{Rogers_2013}%
  \BibitemOpen
  \bibfield  {author} {\bibinfo {author} {\bibfnamefont {E.~T.~F.}\ \bibnamefont {Rogers}}\ and\ \bibinfo {author} {\bibfnamefont {N.~I.}\ \bibnamefont {Zheludev}},\ }\bibfield  {title} {\bibinfo {title} {{Optical super-oscillations: sub-wavelength light focusing and super-resolution imaging}},\ }\href {https://doi.org/10.1088/2040-8978/15/9/094008} {\bibfield  {journal} {\bibinfo  {journal} {J. Optics}\ }\textbf {\bibinfo {volume} {15}},\ \bibinfo {pages} {94008} (\bibinfo {year} {2013})}\BibitemShut {NoStop}%
\bibitem [{\citenamefont {Wong}\ and\ \citenamefont {Eleftheriades}(2011)}]{wong_temporal_2011}%
  \BibitemOpen
  \bibfield  {author} {\bibinfo {author} {\bibfnamefont {A.~M.~H.}\ \bibnamefont {Wong}}\ and\ \bibinfo {author} {\bibfnamefont {G.~V.}\ \bibnamefont {Eleftheriades}},\ }\bibfield  {title} {\bibinfo {title} {Temporal pulse compression beyond the {Fourier} transform limit},\ }\href {https://doi.org/10.1109/TMTT.2011.2160961} {\bibfield  {journal} {\bibinfo  {journal} {IEEE Trans. Microwave Theory Tech.}\ }\textbf {\bibinfo {volume} {59}},\ \bibinfo {pages} {2173} (\bibinfo {year} {2011})}\BibitemShut {NoStop}%
\bibitem [{\citenamefont {Wong}\ and\ \citenamefont {Eleftheriades}(2012)}]{wong_superoscillatory_2012}%
  \BibitemOpen
  \bibfield  {author} {\bibinfo {author} {\bibfnamefont {A.~M.~H.}\ \bibnamefont {Wong}}\ and\ \bibinfo {author} {\bibfnamefont {G.~V.}\ \bibnamefont {Eleftheriades}},\ }\bibfield  {title} {\bibinfo {title} {Superoscillatory radar imaging: Improving radar range resolution beyond fundamental bandwidth limitations},\ }\href {https://doi.org/10.1109/LMWC.2012.2185824} {\bibfield  {journal} {\bibinfo  {journal} {IEEE Microwave Wireless Compon. Lett.}\ }\textbf {\bibinfo {volume} {22}},\ \bibinfo {pages} {147} (\bibinfo {year} {2012})}\BibitemShut {NoStop}%
\bibitem [{\citenamefont {Eliezer}\ \emph {et~al.}(2017)\citenamefont {Eliezer}, \citenamefont {Hareli}, \citenamefont {Lobachinsky}, \citenamefont {Froim},\ and\ \citenamefont {Bahabad}}]{eliezer_breaking_2017}%
  \BibitemOpen
  \bibfield  {author} {\bibinfo {author} {\bibfnamefont {Y.}~\bibnamefont {Eliezer}}, \bibinfo {author} {\bibfnamefont {L.}~\bibnamefont {Hareli}}, \bibinfo {author} {\bibfnamefont {L.}~\bibnamefont {Lobachinsky}}, \bibinfo {author} {\bibfnamefont {S.}~\bibnamefont {Froim}},\ and\ \bibinfo {author} {\bibfnamefont {A.}~\bibnamefont {Bahabad}},\ }\bibfield  {title} {\bibinfo {title} {Breaking the temporal resolution limit by superoscillating optical beats},\ }\href {https://doi.org/10.1103/PhysRevLett.119.043903} {\bibfield  {journal} {\bibinfo  {journal} {Phys. Rev. Lett.}\ }\textbf {\bibinfo {volume} {119}},\ \bibinfo {pages} {043903} (\bibinfo {year} {2017})}\BibitemShut {NoStop}%
\bibitem [{\citenamefont {Eliezer}\ \emph {et~al.}(2018)\citenamefont {Eliezer}, \citenamefont {Singh}, \citenamefont {Hareli}, \citenamefont {Bahabad},\ and\ \citenamefont {Arie}}]{eliezer_experimental_2018}%
  \BibitemOpen
  \bibfield  {author} {\bibinfo {author} {\bibfnamefont {Y.}~\bibnamefont {Eliezer}}, \bibinfo {author} {\bibfnamefont {B.~K.}\ \bibnamefont {Singh}}, \bibinfo {author} {\bibfnamefont {L.}~\bibnamefont {Hareli}}, \bibinfo {author} {\bibfnamefont {A.}~\bibnamefont {Bahabad}},\ and\ \bibinfo {author} {\bibfnamefont {A.}~\bibnamefont {Arie}},\ }\bibfield  {title} {\bibinfo {title} {Experimental realization of structured super-oscillatory pulses},\ }\href {https://doi.org/10.1364/OE.26.004933} {\bibfield  {journal} {\bibinfo  {journal} {Opt. Express}\ }\textbf {\bibinfo {volume} {26}},\ \bibinfo {pages} {4933} (\bibinfo {year} {2018})}\BibitemShut {NoStop}%
\bibitem [{\citenamefont {Brehm}\ \emph {et~al.}(2020)\citenamefont {Brehm}, \citenamefont {Akimov}, \citenamefont {Campion},\ and\ \citenamefont {Kent}}]{brehm_temporal_2020}%
  \BibitemOpen
  \bibfield  {author} {\bibinfo {author} {\bibfnamefont {S.}~\bibnamefont {Brehm}}, \bibinfo {author} {\bibfnamefont {A.~V.}\ \bibnamefont {Akimov}}, \bibinfo {author} {\bibfnamefont {R.~P.}\ \bibnamefont {Campion}},\ and\ \bibinfo {author} {\bibfnamefont {A.~J.}\ \bibnamefont {Kent}},\ }\bibfield  {title} {\bibinfo {title} {Temporal superoscillations of subterahertz coherent acoustic phonons},\ }\href {https://doi.org/10.1103/PhysRevResearch.2.023009} {\bibfield  {journal} {\bibinfo  {journal} {Phys. Rev. Res.}\ }\textbf {\bibinfo {volume} {2}},\ \bibinfo {pages} {023009} (\bibinfo {year} {2020})}\BibitemShut {NoStop}%
\bibitem [{\citenamefont {McCaul}\ \emph {et~al.}(2023)\citenamefont {McCaul}, \citenamefont {Peng}, \citenamefont {Martinez}, \citenamefont {Lindberg}, \citenamefont {Talbayev},\ and\ \citenamefont {Bondar}}]{mccaul_superoscillations_2023}%
  \BibitemOpen
  \bibfield  {author} {\bibinfo {author} {\bibfnamefont {G.}~\bibnamefont {McCaul}}, \bibinfo {author} {\bibfnamefont {P.}~\bibnamefont {Peng}}, \bibinfo {author} {\bibfnamefont {M.~O.}\ \bibnamefont {Martinez}}, \bibinfo {author} {\bibfnamefont {D.~R.}\ \bibnamefont {Lindberg}}, \bibinfo {author} {\bibfnamefont {D.}~\bibnamefont {Talbayev}},\ and\ \bibinfo {author} {\bibfnamefont {D.~I.}\ \bibnamefont {Bondar}},\ }\bibfield  {title} {\bibinfo {title} {Superoscillations {Deliver} {Superspectroscopy}},\ }\href {https://doi.org/10.1103/PhysRevLett.131.153803} {\bibfield  {journal} {\bibinfo  {journal} {Physical Review Letters}\ }\textbf {\bibinfo {volume} {131}},\ \bibinfo {pages} {153803} (\bibinfo {year} {2023})}\BibitemShut {NoStop}%
\bibitem [{\citenamefont {Peng}\ \emph {et~al.}(2025)\citenamefont {Peng}, \citenamefont {Lindberg}, \citenamefont {McCaul}, \citenamefont {Bondar},\ and\ \citenamefont {Talbayev}}]{peng_super-sensing_2025}%
  \BibitemOpen
  \bibfield  {author} {\bibinfo {author} {\bibfnamefont {P.}~\bibnamefont {Peng}}, \bibinfo {author} {\bibfnamefont {D.~R.}\ \bibnamefont {Lindberg}}, \bibinfo {author} {\bibfnamefont {G.}~\bibnamefont {McCaul}}, \bibinfo {author} {\bibfnamefont {D.~I.}\ \bibnamefont {Bondar}},\ and\ \bibinfo {author} {\bibfnamefont {D.}~\bibnamefont {Talbayev}},\ }\bibfield  {title} {\bibinfo {title} {Super-sensing: 100-fold enhancement in {THz} time-domain spectroscopy contrast via superoscillating waveform shaping},\ }\href {https://doi.org/10.1063/5.0271556} {\bibfield  {journal} {\bibinfo  {journal} {APL Photonics}\ }\textbf {\bibinfo {volume} {10}},\ \bibinfo {pages} {086107} (\bibinfo {year} {2025})}\BibitemShut {NoStop}%
\bibitem [{\citenamefont {White}\ \emph {et~al.}(2024)\citenamefont {White}, \citenamefont {Zhang}, \citenamefont {{\v S}oda}, \citenamefont {Kempf}, \citenamefont {Struppa}, \citenamefont {Jordan},\ and\ \citenamefont {Howell}}]{white_reconstructing_2024}%
  \BibitemOpen
  \bibfield  {author} {\bibinfo {author} {\bibfnamefont {D.~D.}\ \bibnamefont {White}}, \bibinfo {author} {\bibfnamefont {S.}~\bibnamefont {Zhang}}, \bibinfo {author} {\bibfnamefont {B.}~\bibnamefont {{\v S}oda}}, \bibinfo {author} {\bibfnamefont {A.}~\bibnamefont {Kempf}}, \bibinfo {author} {\bibfnamefont {D.~C.}\ \bibnamefont {Struppa}}, \bibinfo {author} {\bibfnamefont {A.~N.}\ \bibnamefont {Jordan}},\ and\ \bibinfo {author} {\bibfnamefont {J.~C.}\ \bibnamefont {Howell}},\ }\bibfield  {title} {\bibinfo {title} {Reconstructing superoscillations buried deeply in noise},\ }\href {https://doi.org/10.1103/PhysRevA.110.L061502} {\bibfield  {journal} {\bibinfo  {journal} {Phys. Rev. A}\ }\textbf {\bibinfo {volume} {110}},\ \bibinfo {pages} {L061502} (\bibinfo {year} {2024})}\BibitemShut {NoStop}%
\bibitem [{\citenamefont {Stern}\ \emph {et~al.}(2024)\citenamefont {Stern}, \citenamefont {Bloch}, \citenamefont {Grynszpan}, \citenamefont {Kahn}, \citenamefont {Aharonov}, \citenamefont {Dressel}, \citenamefont {Cohen},\ and\ \citenamefont {Howell}}]{stern_light_2024}%
  \BibitemOpen
  \bibfield  {author} {\bibinfo {author} {\bibfnamefont {I.}~\bibnamefont {Stern}}, \bibinfo {author} {\bibfnamefont {Y.}~\bibnamefont {Bloch}}, \bibinfo {author} {\bibfnamefont {E.}~\bibnamefont {Grynszpan}}, \bibinfo {author} {\bibfnamefont {M.}~\bibnamefont {Kahn}}, \bibinfo {author} {\bibfnamefont {Y.}~\bibnamefont {Aharonov}}, \bibinfo {author} {\bibfnamefont {J.}~\bibnamefont {Dressel}}, \bibinfo {author} {\bibfnamefont {E.}~\bibnamefont {Cohen}},\ and\ \bibinfo {author} {\bibfnamefont {J.~C.}\ \bibnamefont {Howell}},\ }\bibfield  {title} {\bibinfo {title} {Light that appears to come from a source that does not exist},\ }\href {https://doi.org/10.1103/PhysRevA.109.012206} {\bibfield  {journal} {\bibinfo  {journal} {Phys. Rev. A}\ }\textbf {\bibinfo {volume} {109}},\ \bibinfo {pages} {012206} (\bibinfo {year} {2024})}\BibitemShut {NoStop}%
\bibitem [{\citenamefont {Smith}(2007)}]{smith2007mathematics}%
  \BibitemOpen
  \bibfield  {author} {\bibinfo {author} {\bibfnamefont {J.}~\bibnamefont {Smith}},\ }\href@noop {} {\emph {\bibinfo {title} {Mathematics of the discrete Fourier transform (DFT) : with audio applications}}}\ (\bibinfo  {publisher} {BookSurge Publishing (http//www.booksurge.com},\ \bibinfo {address} {Seattle. WA},\ \bibinfo {year} {2007})\BibitemShut {NoStop}%
\bibitem [{\citenamefont {Lee}\ \emph {et~al.}(2000{\natexlab{a}})\citenamefont {Lee}, \citenamefont {Meade}, \citenamefont {Perlin}, \citenamefont {Winful}, \citenamefont {Norris},\ and\ \citenamefont {Galvanauskas}}]{lee_generation_2000}%
  \BibitemOpen
  \bibfield  {author} {\bibinfo {author} {\bibfnamefont {Y.-S.}\ \bibnamefont {Lee}}, \bibinfo {author} {\bibfnamefont {T.}~\bibnamefont {Meade}}, \bibinfo {author} {\bibfnamefont {V.}~\bibnamefont {Perlin}}, \bibinfo {author} {\bibfnamefont {H.}~\bibnamefont {Winful}}, \bibinfo {author} {\bibfnamefont {T.~B.}\ \bibnamefont {Norris}},\ and\ \bibinfo {author} {\bibfnamefont {A.}~\bibnamefont {Galvanauskas}},\ }\bibfield  {title} {\bibinfo {title} {Generation of narrow-band terahertz radiation via optical rectification of femtosecond pulses in periodically poled lithium niobate},\ }\href {https://doi.org/10.1063/1.126390} {\bibfield  {journal} {\bibinfo  {journal} {Applied Physics Letters}\ }\textbf {\bibinfo {volume} {76}},\ \bibinfo {pages} {2505} (\bibinfo {year} {2000}{\natexlab{a}})}\BibitemShut {NoStop}%
\bibitem [{\citenamefont {Lee}\ \emph {et~al.}(2000{\natexlab{b}})\citenamefont {Lee}, \citenamefont {Meade}, \citenamefont {DeCamp}, \citenamefont {Norris},\ and\ \citenamefont {Galvanauskas}}]{lee_temperature_2000}%
  \BibitemOpen
  \bibfield  {author} {\bibinfo {author} {\bibfnamefont {Y.-S.}\ \bibnamefont {Lee}}, \bibinfo {author} {\bibfnamefont {T.}~\bibnamefont {Meade}}, \bibinfo {author} {\bibfnamefont {M.}~\bibnamefont {DeCamp}}, \bibinfo {author} {\bibfnamefont {T.~B.}\ \bibnamefont {Norris}},\ and\ \bibinfo {author} {\bibfnamefont {A.}~\bibnamefont {Galvanauskas}},\ }\bibfield  {title} {\bibinfo {title} {Temperature dependence of narrow-band terahertz generation from periodically poled lithium niobate},\ }\href {https://doi.org/10.1063/1.1290046} {\bibfield  {journal} {\bibinfo  {journal} {Applied Physics Letters}\ }\textbf {\bibinfo {volume} {77}},\ \bibinfo {pages} {1244} (\bibinfo {year} {2000}{\natexlab{b}})}\BibitemShut {NoStop}%
\bibitem [{\citenamefont {Weiss}\ \emph {et~al.}(2001)\citenamefont {Weiss}, \citenamefont {Torosyan}, \citenamefont {Meyn}, \citenamefont {Wallenstein}, \citenamefont {Beigang},\ and\ \citenamefont {Avetisyan}}]{weiss_tuning_2001}%
  \BibitemOpen
  \bibfield  {author} {\bibinfo {author} {\bibfnamefont {C.}~\bibnamefont {Weiss}}, \bibinfo {author} {\bibfnamefont {G.}~\bibnamefont {Torosyan}}, \bibinfo {author} {\bibfnamefont {J.-P.}\ \bibnamefont {Meyn}}, \bibinfo {author} {\bibfnamefont {R.}~\bibnamefont {Wallenstein}}, \bibinfo {author} {\bibfnamefont {R.}~\bibnamefont {Beigang}},\ and\ \bibinfo {author} {\bibfnamefont {Y.}~\bibnamefont {Avetisyan}},\ }\bibfield  {title} {\bibinfo {title} {Tuning characteristics of narrowband {THz} radiation generated via optical rectification in periodically poled lithium niobate},\ }\href {https://doi.org/10.1364/OE.8.000497} {\bibfield  {journal} {\bibinfo  {journal} {Optics Express}\ }\textbf {\bibinfo {volume} {8}},\ \bibinfo {pages} {497} (\bibinfo {year} {2001})},\ \bibinfo {note} {publisher: Optica Publishing Group}\BibitemShut {NoStop}%
\bibitem [{\citenamefont {L’huillier}\ \emph {et~al.}(2007{\natexlab{a}})\citenamefont {L’huillier}, \citenamefont {Torosyan}, \citenamefont {Theuer}, \citenamefont {Avetisyan},\ and\ \citenamefont {Beigang}}]{lhuillier_generation_2007}%
  \BibitemOpen
  \bibfield  {author} {\bibinfo {author} {\bibfnamefont {J.}~\bibnamefont {L’huillier}}, \bibinfo {author} {\bibfnamefont {G.}~\bibnamefont {Torosyan}}, \bibinfo {author} {\bibfnamefont {M.}~\bibnamefont {Theuer}}, \bibinfo {author} {\bibfnamefont {Y.}~\bibnamefont {Avetisyan}},\ and\ \bibinfo {author} {\bibfnamefont {R.}~\bibnamefont {Beigang}},\ }\bibfield  {title} {\bibinfo {title} {Generation of {THz} radiation using bulk, periodically and aperiodically poled lithium niobate – {Part} 1: {Theory}},\ }\href {https://doi.org/10.1007/s00340-006-2490-9} {\bibfield  {journal} {\bibinfo  {journal} {Applied Physics B}\ }\textbf {\bibinfo {volume} {86}},\ \bibinfo {pages} {185} (\bibinfo {year} {2007}{\natexlab{a}})}\BibitemShut {NoStop}%
\bibitem [{\citenamefont {L’huillier}\ \emph {et~al.}(2007{\natexlab{b}})\citenamefont {L’huillier}, \citenamefont {Torosyan}, \citenamefont {Theuer}, \citenamefont {Rau}, \citenamefont {Avetisyan},\ and\ \citenamefont {Beigang}}]{lhuillier_generation_2007-1}%
  \BibitemOpen
  \bibfield  {author} {\bibinfo {author} {\bibfnamefont {J.}~\bibnamefont {L’huillier}}, \bibinfo {author} {\bibfnamefont {G.}~\bibnamefont {Torosyan}}, \bibinfo {author} {\bibfnamefont {M.}~\bibnamefont {Theuer}}, \bibinfo {author} {\bibfnamefont {C.}~\bibnamefont {Rau}}, \bibinfo {author} {\bibfnamefont {Y.}~\bibnamefont {Avetisyan}},\ and\ \bibinfo {author} {\bibfnamefont {R.}~\bibnamefont {Beigang}},\ }\bibfield  {title} {\bibinfo {title} {Generation of {THz} radiation using bulk, periodically and aperiodically poled lithium niobate – {Part} 2: {Experiments}},\ }\href {https://doi.org/10.1007/s00340-006-2528-z} {\bibfield  {journal} {\bibinfo  {journal} {Applied Physics B}\ }\textbf {\bibinfo {volume} {86}},\ \bibinfo {pages} {197} (\bibinfo {year} {2007}{\natexlab{b}})}\BibitemShut {NoStop}%
\bibitem [{\citenamefont {Lin}\ \emph {et~al.}(2018)\citenamefont {Lin}, \citenamefont {Yu},\ and\ \citenamefont {Talbayev}}]{lin_measurement_2018}%
  \BibitemOpen
  \bibfield  {author} {\bibinfo {author} {\bibfnamefont {S.}~\bibnamefont {Lin}}, \bibinfo {author} {\bibfnamefont {S.}~\bibnamefont {Yu}},\ and\ \bibinfo {author} {\bibfnamefont {D.}~\bibnamefont {Talbayev}},\ }\bibfield  {title} {\bibinfo {title} {Measurement of {Quadratic} {Terahertz} {Optical} {Nonlinearities} {Using} {Second}-{Harmonic} {Lock}-in {Detection}},\ }\href {https://doi.org/10.1103/PhysRevApplied.10.044007} {\bibfield  {journal} {\bibinfo  {journal} {Physical Review Applied}\ }\textbf {\bibinfo {volume} {10}},\ \bibinfo {pages} {044007} (\bibinfo {year} {2018})},\ \bibinfo {note} {publisher: American Physical Society}\BibitemShut {NoStop}%
\bibitem [{\citenamefont {Bondar}\ and\ \citenamefont {Talbayev}(2026)}]{your_code_ref}%
  \BibitemOpen
  \bibfield  {author} {\bibinfo {author} {\bibfnamefont {D.~I.}\ \bibnamefont {Bondar}}\ and\ \bibinfo {author} {\bibfnamefont {D.}~\bibnamefont {Talbayev}},\ }\href {https://github.com/dibondar/bicolor-superoscillations} {\bibinfo {title} {Code and data for synthesizing bichromatic superoscillations}} (\bibinfo {year} {2026}),\ \bibinfo {note} {gitHub repository}\BibitemShut {NoStop}%
\end{thebibliography}%

\end{document}